\documentclass[english]{article}
\usepackage[T1]{fontenc}
\usepackage[latin9]{inputenc}
\usepackage{graphicx}

\makeatletter

\newcommand{\lyxaddress}[1]{
\par {\raggedright #1
\vspace{1.4em}
\noindent\par}
}

\makeatother

\makeatother

\usepackage{babel}

\begin{document}

\title{A causality analysis of the linearized relativistic Navier-Stokes
equations}

\author{A. Sandoval-Villalbazo$^{1}$, A. L. Garcia-Perciante$^{2}$}

\maketitle

\lyxaddress{$^{1}$Depto. de Fisica y Matematicas, Universidad Iberoamericana,
Prolongacion Paseo de la Reforma 880, Mexico D. F. 01210, Mexico.\\
 $^{2}$Depto. de Matematicas Aplicadas y Sistemas, Universidad
Autonoma Metropolitana-Cuajimalpa, Artificios 40 Mexico D.F 01120,
Mexico. }
\begin{abstract}
It is shown by means of a simple analysis that the linearized system
of transport equations for a relativistic, single component ideal
gas at rest obeys the \textit{antecedence principle}, which is often
referred to as causality principle. This task is accomplished by examining
the roots of the dispersion relation for such a system. This result
is important for recent experiments performed in relativistic heavy
ion colliders, since it suggests that the Israel-Stewart like formalisms
may be unnecessary in order to describe relativistic fluids.
\end{abstract}

\section{Introduction}

In 1940 C. Eckart published three papers entitled \emph{The thermodynamics
of irreversible processes} \cite{eckart1}, the third one addressing
the problem of a relativistic simple (single component ideal gas)
fluid. In that paper, Eckart proceeded following the basic ideas of
classical irreversible thermodynamics \cite{Groot1}, except for the
fact that he introduced relativistic terms in the energy-momentum
tensor. As part of his phenomenological approach, he proposed constitutive
equations with relativistic corrections. Since Eckart's theory apparently
leads to results that violate causality and involves undesirable unstable
modes \cite{Hiscock}, it has been patched up in several ways using
formalisms introduced by Israel and coworkers \cite{Israel1} \cite{Israel2}
\cite{Israel3} and sometimes using extended irreversible thermodynamics
\cite{Zimdahl} \cite{Jou4}. Recently, it has been shown that the
unphysical behavior of the unstable modes is due to the coupling between
heat an acceleration proposed by Eckart \cite{GRG09}. Indeed, it
has been shown that such a relation is not sustained by kinetic theory
\cite{PA09}.

The so-called \textit{causality problem} of heat conduction, which
should be more precisely stated as \emph{antecedence problem}, remains
still a controversial issue which suggests the need of extended theories.
We wish to point out that eventhough the term causality is the most
favoured, the problem of faster that light propagation of fluctuations
is not striclty a cause-effect issue but rather the prediction of
an unphysical behavior concerning arrival times. However, as is shown
in the following sections, relativistic classical linear irreversible
thermodynamics, as obtained from relativistic kinetic theory features
both stability for the equilibrium state \cite{PA09} and satisfaction
of the antecedence principle.

To accomplish this task we divide the rest of the paper as follows.
In Sect. 1.2 we recall the Navier-Stokes equations for a simple relativistic
fluid \cite{pre09} and introduce the appropriate constitutive equation
for the heat flux. The linearized set of transport equations is thoroughly
analyzed in Sect. 1.3. Conclusions and final remarks are included
in Sect. 1.4.

\section{Transport equations for the relativistic single component ideal gas}

The starting point are the balance equations for a relativistic fluid
which are obtained from the conservation of the particle density flow\begin{equation}
N^{\nu}=nu^{\nu}\label{eq:1}\end{equation}
 and the energy-momentum tensor which, following Eckart\cite{eckart1}
reads\begin{equation}
T_{\nu}^{\mu}=\frac{n\varepsilon}{c^{2}}u^{\mu}u_{\nu}+ph_{\nu}^{\mu}+\pi_{\nu}^{\mu}+\frac{1}{c^{2}}q^{\mu}u_{\nu}+\frac{1}{c^{2}}u^{\mu}q_{\nu}\label{eq:2}\end{equation}
In Eqs. (\ref{eq:1}) and (\ref{eq:2}), $n$ is the particle number
density, $u^{\nu}$ the hydrodynamic velocity four vector, $c$ the
speed of light, $p$ the hydrostatic pressure and $h_{\nu}^{\mu}=\delta_{\nu}^{\mu}+u^{\mu}u_{\nu}/c^{2}$
the spatial projector. The internal energy per particle, $\varepsilon$
, includes the rest energy since it is given by \cite{cermed}\begin{equation}
\varepsilon=mc^{2}\left(3z+\frac{\mathcal{K}_{1}\left(\frac{1}{z}\right)}{\mathcal{K}_{2}\left(\frac{1}{z}\right)}\right)\sim mc^{2}+\frac{3}{2}kT+...\label{eq:mass}\end{equation}
where $z=\frac{kT}{mc^{2}}$ is the relativistic parameter and $\mathcal{K}_{n}\left(\frac{1}{z}\right)$
are the modified Bessel function of the second kind. The dissipative
fluxes are the Navier tensor $\pi_{\nu}^{\mu}$ and the heat flux
$q^{\nu}$ . The conservation equations $N_{;\nu}^{\nu}=0$ and $T_{\nu;\mu}^{\mu}=0$
for the quantities defined above yield the Navier-Stokes equations
for the relativistic simple fluid namely,

\begin{equation}
\dot{n}+n\theta=0\label{eq:3}\end{equation}
 \begin{eqnarray}
\left(\frac{n\varepsilon}{c^{2}}+\frac{p}{c^{2}}\right)\dot{u}_{\nu}+\left(\frac{n\dot{\varepsilon}}{c^{2}}+\frac{p}{c^{2}}\theta\right)u_{\nu}+p_{,\mu}h_{\nu}^{\mu}+\pi_{\nu,\mu}^{\mu}\nonumber \\
+\frac{1}{c^{2}}\left(q_{,\mu}^{\mu}u_{\nu}+q^{\mu}u_{\nu,\mu}+\theta q_{\nu}+u^{\mu}q_{\nu,\mu}\right)=0\label{eq:4}\end{eqnarray}

\begin{equation}
nC_{n}\dot{T}+\left(\frac{T\beta}{\kappa_{T}}\right)\theta+u_{,\mu}^{\nu}\pi_{\nu}^{\mu}+q_{;\mu}^{\mu}+\frac{1}{c^{2}}\dot{u}^{\nu}q_{\nu}=0\label{eq:5}\end{equation}
where $\kappa_{T}$ is the isothermal compressibility, $\beta$ the
thermal expansion coefficient and $C_{n}$ the heat capacity at constant
particle density. As it will be remarked below, the tensor $\pi_{\nu}^{\mu}$
can be further decomposed in a traceless symmetric part and the trace
multiplied by the spatial projector.

In order to close the system of equations, constitutive relations
for the heat flux and Navier tensor must be introduced. The equation
for the heat flux has been recently established by means relativistic
kinetic theory and reads{\small \begin{equation}
q^{\ell}=-L_{T}\frac{T^{,\ell}}{T}+L_{n}\frac{n^{,\ell}}{n}\label{eq:6}\end{equation}
where $L_{T}$ and $L_{n}$ are transport coefficients\cite{PA09}.
A detailed discussion on this equations can be found elsewhere\cite{PA09}.
The equations for $\pi_{\nu}^{\mu}$ in Eq. (\ref{eq:2}) are well-known
namely,\begin{equation}
\pi_{\mu\nu}^{\left(s\right)}=-2\eta\sigma_{\mu\nu}\label{eq:6.1}\end{equation}
 \begin{equation}
tr\left(\pi\right)=-\xi\nabla\cdot\vec{u}\label{eq:6.2}\end{equation}
 where $\pi_{\mu\nu}^{\left(s\right)}$ is the symmetric and traceless
part of $\pi_{\nu}^{\mu}$, $tr\left(\pi\right)$ its trace and $\sigma_{\mu\nu}$
is the symmetric and taceless part of the velocity gradient. The transport
coefficients in Eqs. (\ref{eq:6.1}) and (\ref{eq:6.2}) are the shear
and bulk viscosities respectively.}{\small \par}

\section{Linearized relativistic hydrodynamics}

In order to linearize the set of equations (\ref{eq:3}-\ref{eq:5})
we consider $n=n_{0}+\delta n$, $T=T_{0}+\delta T$ and $u^{\nu}=\delta u^{\nu}$
where naught subscripts denote equilibrium quantities and the $\delta$
prefix indicates small perturbations around it. With this hypothesis,
the linearized transport equations for a simple, relativistic fluid
in the absence of external fields are \begin{equation}
\delta\dot{n}+n_{0}\delta\theta=0\label{eq:7}\end{equation}
 \begin{eqnarray}
\frac{1}{c^{2}}\left(n_{0}\varepsilon_{0}+p_{0}\right)\delta\dot{u}_{\nu}+\frac{1}{n\kappa_{T}}\delta n_{,\nu}+\frac{\beta}{\kappa_{T}}\delta T_{,\nu}\nonumber \\
-\zeta\delta\theta_{,\nu}-2\eta\left(\delta\sigma_{\nu}^{\mu}\right)_{,\mu}-\frac{L_{T}}{c^{2}}\delta\dot{T}_{,\nu}-\frac{L_{n}}{c^{2}}\delta\dot{n}_{,\nu} & =0\label{eq:8}\end{eqnarray}
 \begin{equation}
nC_{n}\delta\dot{T}+\left(\frac{T_{0}\beta}{\kappa_{T}}\right)\delta\theta-\left(L_{T}\delta T^{,k}+L_{n}\delta n^{,k}\right)_{;k}=0\label{eq:9}\end{equation}
 where we have defined $\theta=u_{;\nu}^{\nu}$. It is important to
point out that the transport coefficients in general depend on the
state variables. However, since they only appear as factors of derivatives
of the corresponding fluctuations, considering fluctuations on them
would induce higher order terms, which are neglected in the linear
approximation.

It is crucial at this point to make the following observation. The
so-called causality violation of the transport equations to first
order in the gradients, given by linear irreversible thermodynamics,
can be easily spotted by observing that, considering $u_{0}^{\ell}=0$
and linearizing, Eq. (\ref{eq:5}) leads to a parabolic equation for
$T$. This clearly admits arbitrary propagation speeds for the corresponding
signals. However, the hypothesis of a fluid at rest or the fact that
calculations can be performed in the comoving frame should not be
translated into a vanishing hydrodynamic velocity, but in $u_{0}^{\ell}=0$
as considered above. That is, $\delta u^{\nu}$ should not vanish
even for the fluid at rest or in the comoving frame; \textit{only
the mean or equilibrium velocity can be zero. }This fact has already
been pointed out in the analysis of the linearized relativistic Euler
regime\cite{domi}.

The analysis of the dynamics given the system of equations (\ref{eq:7}-\ref{eq:9})
can be found in detail in Section 4 of Ref. \cite{pre09} where we
discussed the modifications to the Rayleigh-Brillouin spectrum. Here
we only quote the results needed in order to address the problem at
hand namely, the \emph{causality} of the system. We start by calculating
the divergence of Eq. (\ref{eq:8}). The transverse mode is then uncoupled
from the system and a set of three scalar differential equations for
$\delta n$, $\delta\theta$ and $\delta T$ is obtained. A Fourier-Laplace
transform is then performed, leading to a system of algebraic equations
depending on the time and space variables, $s$ and $q$ respectively,
whose associated determinant reads\begin{equation}
\left|\begin{array}{ccc}
s & n_{0} & 0\\
-\frac{1}{n_{0}\kappa_{T}}q^{2}+\frac{L_{n}}{c^{2}}sq^{2} & \tilde{\rho}_{0}s+Aq^{2} & \frac{L_{T}}{c^{2}}q^{2}s-\frac{\beta}{\kappa_{T}}q^{2}\\
\frac{L_{n}}{n_{0}c_{n}}q^{2} & \frac{T_{0}\beta}{n_{0}c_{n}\kappa_{T}} & s+\frac{L_{TT}}{n_{0}c_{n}}q^{2}\end{array}\right|=0\label{eq:10}\end{equation}
where for convenience we have introduced the following notation\cite{pre09}\begin{equation}
\tilde{\rho}_{0}=\frac{1}{c^{2}}\left(n_{0}\varepsilon_{0}+p_{0}\right)\label{eq:11}\end{equation}
 \begin{equation}
A=\zeta+4\eta/3\label{eq:12}\end{equation}

\noindent The dispersion relation is thus given by\begin{equation}
s^{3}+d_{2}s^{2}q^{2}+s\left(d_{3}q^{4}+d_{4}q^{2}\right)+d_{5}q^{4}=0\label{eq:fr5}\end{equation}
where the coefficients $d_{2}$ to $d_{5}$ have been specified in
an earlier work\cite{pre09}. The physical interpretation of the three
roots of Eq. (\ref{eq:fr5}) is well known. The dynamics of the perturbations
in the fluid are characterized by a strictly dissipative component
which decays in time depending on the value of the real root while
the other wave-like component propagates at a speed given by the imaginary
parts of the conjugate roots, damped by a coefficient which depends
on a Stokes-Kirchhoff like factor. Moreover, a plot of the dynamic
structure factor as a function of $s$ for a fixed $\vec{q}$, will
feature three peaks. In this work, we are interested in the location
of the symmetric Brillouin peaks\cite{berne}, which are given by
the imaginary part of the conjugate roots, that is $\omega=\pm\sqrt{d_{4}}q$.
Thus, in this case \begin{equation}
\omega=\pm\sqrt{\frac{\gamma}{\kappa_{T}\tilde{\rho}_{0}}}q\label{eq:as}\end{equation}
 such that, the distance between the peaks, i. e. the speed of propagation
of the wave-like component of the fluctuations, and the origin is
bounded by\begin{equation}
c\sqrt{\frac{\gamma}{\kappa_{T}\left(n_{0}\varepsilon_{0}+p_{0}\right)}}\label{eq:13}\end{equation}
 Notice that, in the non-relativistic case, the fluctuations propagate
at the speed of sound, i. e.\begin{equation}
c_{s}^{2}=\frac{\gamma}{\kappa_{T}\rho_{0}}\label{eq:14}\end{equation}
As an example, for an ideal gas $\gamma=5/3$ and $\kappa_{T}=1/p$,
such that\begin{equation}
c_{s}^{2}=\frac{5}{3}\frac{kT}{m}\label{eq:15}\end{equation}
which is clearly unbounded and can be increasingly large for high
temperatures. However, the speed of propagation in the relativistic
calculation reads\begin{equation}
c_{R}^{2}=\frac{\gamma}{\kappa_{T}\left(n_{0}\varepsilon_{0}+p_{0}\right)}c^{2}\label{eq:16}\end{equation}
Using the expression for the internal energy density given by Eq.
(\ref{eq:mass}) we now obtain\begin{equation}
c_{R}^{2}=\left\{ \frac{5}{3}\frac{z}{\left[3z+\frac{\mathcal{K}_{1}\left(\frac{1}{z}\right)}{\mathcal{K}_{2}\left(\frac{1}{z}\right)}\right]+1}\right\} c^{2}\label{eq:18}\end{equation}

\noindent As can be seen in Fig. 1, the expression in curly brackets
never exceeds the unity which finally shows that the propagation speed
for signals in the transport equations for the relativistic fluid
does not violate the antecedence principle.
\begin{figure}

\caption{The ratio of the speed of propagation to the speed of light, as a
function of the relativistic parameter $z$.}
\begin{centering}
\includegraphics[
  width=4in,
  height=2.5in]{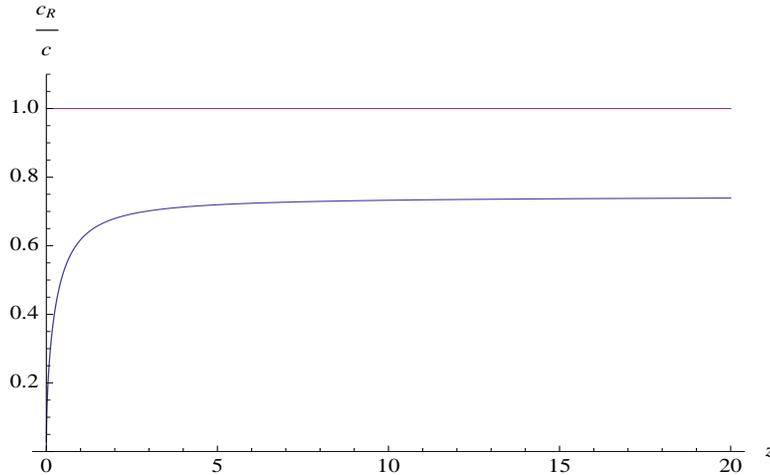}
\end{centering}
\label{fig:F1}

\end{figure}

\section{Final remarks}

The transport equations derived from Eckart's modified theory show,
in its linear version, no problems regarding stability and causality
features. This fact implies that there is no real motivation to introduce
the so-called second order theories, which introduce non-fundamental
adjustable parameters. A simple first order formalism is desirable
to describe the fluids which are formed in RHIC type experiments\cite{wong}.

In the non-linear case the relativistic Navier-Stokes equations here
employed are, by far, more complex. In this context, very little can
be said regarding the problems of stability and causality with the
techniques included in this paper. It is desirable to perform further
work in this direction.

\end{document}